\documentclass[preprintnumbers,amsmath,11pt,amssymb,floatfix,superscriptaddress,nofootinbib]{article}

\def\mytitle#1{\setcounter{equation}{0}
\setcounter{footnote}{0}
\begin{flushleft}\Large\textbf{#1}\end{flushleft}
\vspace{0.25cm}}
\def\myname#1{\leftline{{\large #1}}\vspace{-0.13cm}}
\def\myplace#1#2{\small\begin{flushleft}\textit{#1}\\
\texttt{#2}\end{flushleft}}

\def\myclassification#1{\small\noindent
Pacs no :
       #1\vspace{0.5cm}}
\usepackage{graphicx}
\begin{document}

\mytitle{Nature of Singularity Formed by the Gravitational
Collapse in Husain Space-Time with Electromagnetic Field and
Scalar Field}

\myname{ Ujjal Debnath~\footnote{ujjal@iucaa.ernet.in,
ujjaldebnath@yahoo.com}$*$} \vskip0.2cm \myname{ Prabir
Rudra~\footnote{prudra.math@gmail.com}$*$} \vskip0.2cm \myname{
Ritabrata Biswas~\footnote{biswas.ritabrata@gmail.com}$\dag$}

\myplace{$*$Department of Mathematics, Bengal Engineering and
Science University, Shibpur, Howrah-711 103, India.} {}

\myplace{$\dag$  Department of Mathematics, Jadavpur University,
Kolkata-700 032, India.} {}

\myclassification{04.20.Dw, 04.20.Ex, 04.20.Cv, 04.70.Bw}

\begin{abstract}
In this work, we have investigated the outcome of gravitational
collapse in Husain space-time in the presence of electro-magnetic
and a scalar field with potential. In order to study the nature of
the singularity, global behavior of radial null geodesics have
been taken into account. The nature of singularities formed has
been thoroughly studied for all possible variations of the
parameters. These choices of parameters has been presented in
tabular form in various dimensions. It is seen that irrespective
of whatever values of the parameters chosen, the collapse always
results in a naked singularity in all dimensions. There is less
possibility of formation of a black hole. Hence this work is a
significant counterexample of the cosmic censorship hypothesis.
\end{abstract}

\section{\normalsize\bf{Introduction}}

When a massive star is on the verge of completing its nuclear
cycle, then the thermonuclear reactions in the interior of the
star cannot counter balance the immense gravitational pull of the
star. Under most general conditions general relativity predicts
that such a collapse must end in a singularity, which may or may
not be clothed by an event horizon. A singularity may be
physically described as a region in the space-time with extreme
curvature, vanishing volume and unbounded gravitational forces.
However, general relativity remains silent on the nature (BH or
NS) or physical properties of such a singularity. This is
basically due to the fact that mathematical structure breaks down
preventing analysis at and beyond the singularity. This has
triggered extensive research on Gravitational collapse during the
past few decades. After all one would always like to know whether,
and under what conditions gravitational collapse leads to the
formation of a black hole (BH). A few decades back R. Penrose
(1969) proposed the cosmic censorship hypothesis (CCH), which
states that the singularities formed in gravitational collapse of
physically reasonable matter cannot be seen by any distant
observer in the universe. It implies that the singularities formed
in asymptotically flat space-times are always bounded by event
horizons and hence are destined to be black holes. With the
announcement of this proposal, study of gravitational collapse has
gained special importance, because one would always like to know
that whether there exists any physical collapse solutions that
lead to naked singularities (NS), which will serve as
counterexamples of CCH.

Till date there has been a lot of significant work in
gravitational collapse. The pioneering work of gravitational
collapse appeared in the famous paper of Oppenheimer and Snyder
(1939) in which they studied the gravitational
collapse of adiabatically flowing dust. The outcome of the
experiment led to the conclusion that the end state of collapse is
a BH. Since then extensive research has been carried out to find
more generalized results. But it took no less than 30 years since
the Oppenheimer-Snyder model that the feasibility of NS was
brought out by Christodoulou (1984) in his study of
Lemaitre-Tolman-Bondi (LTB) model. It is thus understandable that,
due to a lack of any alternative scenario all these years, the
black hole became the unique end state of continued gravitational
collapse for a remnant mass of a collapsing star beyond the
threshold neutron star mass limit. The absence of analytical
results led to several conjectures as well, namely, the (weak and
strong) cosmic censorship conjecture (CCC) by Penrose, hoop
conjecture (HC) by Thorne (1972), and Seifert
(1979) conjecture, which to date remain unproven.

In order to study gravitational collapse properly, it is necessary
to describe adequately the geometry of interior and exterior
regions and to give conditions which allow the matching of these
regions. In this context it is worthy to mention that collapse
with static exterior was studied by Misner and Sharp (1964) for a perfect fluid in the interior. The concept of
outgoing radiation in order to study the outcome of gravitational
collapse was first used by Vaidya (1951). Later on Santos
and collaborations (Santos, N.O. 1984; Oliveira, A.K.G. 1985,1987), Ghosh and
Deskar (2000,2003) and Cissoko et al (1998)
have done meaningful work in gravitational collapse, thus
extending the topic to new heights. The credit of studying
gravitational collapse with a model of arbitrary shape
(non-spherical) goes to Szekeres (1975) who in his model (known as
Szekeres' model with perfect fluid (or dust)
gave detailed solution of gravitational collapse for
quasi-spherical or quasi-cylindrical symmetry of the space-time.
Debnath et al (2005) explored gravitational collapse of
the non-adiabatic fluid by assuming quasi-spherical Szekeres
space-time in the interior and plane symmetric Vaidya solution in
the exterior region. By using the local conservation of momentum,
they studied the thermodynamical behavior of the collapsing
matter. An extensive survey by Herrera (2004) and Mitra
(2006) predicts that gravitational collapse is highly
dissipative process. This showed that the effects of dissipation
must be included in the study of gravitational collapse for its
better understanding. Herrera and Santos (2004) then
explored dynamical description of gravitational collapse by using
Misner and Sharps formulation. Nolan (2002) investigated
naked singularities in the cylindrical gravitational collapse of
counter rotating dust shell. Di Prisco (2009) discussed
the shear free cylindrical gravitational collapse by using
junction conditions. Nakao (2009) studied gravitational
collapse of a hollow cylinder composed of dust. Recently, Sharif
and Rehmat (2010) discussed the dynamics of viscous
dissipative plane symmetric gravitational collapse.

The general problem of gravitational collapse always remains
intractable due to the complex and tedious form of the Einstein
equations. Therefore instead of the general form of the equations,
special forms are only considered to reach meaningful results. Two
dimensional reduction of general relativity using spherical
symmetry is one such case. Walking on this very path Husain
in 1996, gave non-static spherically symmetric solutions of the
Einstein equations for a null fluid source with pressure $p$ and
density $\rho$ related by $p=k\rho$. The solution has a number of
interesting features including BHs with short hair (Brown,J.D. and Husain,V. 1997).
This solution is considered as a generalization of Vaidya
solution, which was accomplished by Wang,A. and Wu,Y. (1999). Recently,
Patil et al (2005,2006) have studied the gravitational collapse of
the Husain solution in four and five dimensional space-times.

In this paper we mainly propose to investigate the nature of
singularity (BH or NS) formed by the collapse of a star in Husain
metric accompanied by a electro-magnetic and a scalar field. In
section (\ref{chap1}) we present the basic equations for Husain
solution in $(n+2)$ dimensional spherically symmetric space-time
coupled with an electro-magnetic and a scalar field and we
investigate the nature of singularity formed. We study the effect
of accretion phenomenon on collapsing procedure in section
(\ref{chap2}). Finally, we end with some concluding remarks in
section (\ref{chap3}).
\section{Basic Equations for Husain Solution in $(n+2)$-Dimensional Spherically
Symmetric Space-Time with Electro-Magnetic Field and Scalar Field}\label{chap1}
Here we consider the metric in $(n+2)$-dimensional spherically
symmetric space-time in the form
\begin{equation}\label{collapse2.1}
ds^{2}=-\left(1-\frac{m(v,r)}{r^{n-1}}\right)dv^{2}+2dvdr+r^{2}d\Omega_{n}^{2}
\end{equation}
where $r$ is the radial co-ordinate and $v$ is the null
co-ordinate, $m(v,r)$ gives the gravitational mass inside the
sphere of radius $r$ and $d\Omega_{n}^{2}$ is the line element on
a unit $n$-sphere.

The Einstein's field equation for matter field, electro-magnetic
field and scalar field with cosmological constant ($\Lambda$) has
the following form (choosing $8\pi G=c=1$)
\begin{equation}\label{collapse2.2}
G_{\mu\nu}=T_{\mu\nu}+E_{\mu\nu}+H_{\mu\nu}-\Lambda g_{\mu\nu}
\end{equation}
Now we consider two types of fluids like Vaidya null radiation and
a perfect fluid having the form of the energy momentum tensor
\begin{equation}\label{collapse2.3}
T_{\mu\nu}=T_{\mu\nu}^{(n)}+T_{\mu\nu}^{(m)}
\end{equation}
with
\begin{equation}\label{collapse2.4}
T_{\mu\nu}^{(n)}=\sigma l_{\mu}l_{\nu}
\end{equation}
and
\begin{equation}\label{collapse2.5}
T_{\mu\nu}^{(m)}=(\rho+p)(l_{\mu}\eta_{\nu}+l_{\nu}\eta_{\mu})+pg_{\mu\nu}
\end{equation}

Where, $\rho$ and $p$ are the energy density and pressure for the
perfect fluid and $\sigma$ is the energy density corresponding to
Vaidya null radiation. In the co-moving co-ordinates
($v,r,\theta_{1},\theta_{2},...,\theta_{n}$), the two eigen
vectors of energy-momentum tensor namely $l_{\mu}$ and
$\eta_{\mu}$ are linearly independent future pointing null vectors
having components
\begin{equation}\label{collapse2.6}
l_{\mu}=(1,0,0,...,0)~ and~
\eta_{\mu}=\left(\frac{1}{2}\left(1-\frac{m}{r^{n-1}}\right),-1,0,...,0
\right)
\end{equation}
and they satisfy the relations
\begin{equation}\label{collapse2.7}
l_{\lambda}l^{\lambda}=\eta_{\lambda}\eta^{\lambda}=0,~ l_{\lambda}\eta^{\lambda}=-1
\end{equation}

The energy- momentum tensor ~$E_{\mu\nu}$~ of the electro-magnetic
field is given by
\begin{equation}\label{collapse2.8}
E_{\mu\nu}=\frac{1}{4\pi}\left[F^{\nu}_{\mu}~F_{\mu\nu}-\frac{1}{4}g_{\mu\nu}F^{\nu\delta}F_{\nu\delta}\right]
\end{equation}
where ~$F_{\mu\nu}$~ is the electro-magnetic field tensor
satisfying Maxwell's equations
\begin{equation}\label{collapse2.9}
F_{\mu\nu}=\psi_{\nu,\mu}-\psi_{\mu,\nu}
\end{equation}
and
\begin{equation}\label{collapse2.10}
\frac{1}{\sqrt{-g}}~\partial_{\mu}\left(\sqrt{-g}~
F^{\mu\nu}\right)=-4\pi J^{\nu}
\end{equation}
where ~$\psi_{\mu}$~ is the four-potential and ~$J_{\mu}$~ is the
four-current.

Since the charge is at rest in this system, so there will be no
magnetic field in this system. Thus, we may choose the
four-potential and four-current as
\begin{equation}\label{collapse2.11}
\psi_{\mu}=\left(\psi(v,r), 0, 0, 0\right)
\end{equation}
and
\begin{equation}\label{collapse2.12}
J^{\mu}=\tau~l^{\mu}
\end{equation}
where ~$\tau$~ is the charge density.

The energy-momentum tensor $H_{\mu\nu}$ for the scalar field $\phi(v,r)$ driven by the potential $V(\phi)$ is given by
\begin{equation}\label{collapse2.13}
H_{\mu\nu}=\phi,_{\mu}\phi,_{\nu}-\frac{1}{2}~g_{\mu\nu}g^{\alpha\beta}\phi,_{\alpha}\phi,_{\beta}-g_{\mu\nu}V(\phi)
\end{equation}
with
\begin{equation}\label{collapse2.14}
\frac{1}{\sqrt{-g}}~\partial_{\mu}\left(\sqrt{-g}~ \partial^{\mu}\phi\right)=-\frac{dV(\phi)}{d\phi}
\end{equation}
From equations (\ref{collapse2.9}) - (\ref{collapse2.12}), we have the only non-zero component of
the field tensor as
\begin{equation}\label{collapse2.15}
F_{01}=-F_{10}=-\psi'=-\frac{e(v)}{r^{n}}
\end{equation}
where $e(v)$ is the arbitrary function of $v$.

From Einstein's field equation (\ref{collapse2.2}), we have
\begin{equation}\label{collapse2.16}
\sigma=\frac{n\dot{m}}{2r^{n}}-\dot{\phi}^{2}
\end{equation}
\begin{equation}\label{collapse2.17}
\rho=\frac{nm'}{2 r^{n}}-\frac{e^{2}(v)}{8\pi
r^{2n}}-V(\phi)-\Lambda
\end{equation}
\begin{equation}\label{collapse2.18}
p=-\frac{m''}{2 r^{n-1}}-\frac{e^{2}(v)}{8\pi
r^{2n}}+V(\phi)+\Lambda
\end{equation}
\begin{equation}\label{collapse2.19}
V(\phi)=\frac{E(v)}{r}+D(r)
\end{equation}
where an over-dot and dash stand for differentiation with respect
to $v$ and $r$ respectively and $E(v)=-\int\dot{\phi}^{2}dv$ and
$D(r)$ is an arbitrary integration function.

We assume the matter fluid obeys the barotropic equation of state
\begin{equation}\label{collapse2.20}
p=k\rho,~~~(k,~a~constant)
\end{equation}
Now using (\ref{collapse2.17}), (\ref{collapse2.18}) and (20) in
the Einstein field equations, we have the differential equation in
$m$,
\begin{equation}\label{collapse2.21}
r^{2}m''+knrm'=\frac{(k-1)e^{2}(v)}{4\pi
r^{n-1}}+2(k+1)\left[\frac{E(v)}{r}+D(r)+\Lambda \right]r^{n+1}
\end{equation}
Let us choose $D(r)=D_{0}r^{\beta}$, for which the explicit
solution for $m$ is ($nk\ne 1$)
\begin{equation}\label{collapse2.22}
m(v,r)=f(v)-\frac{g(v)}{(nk-1)r^{nk-1}}-\frac{e^{2}(v)}{4\pi
n(n-1)r^{n-1}}+\frac{2\Lambda
r^{n+1}}{n(n+1)}+\frac{2(k+1)E(v)r^{n}}{n(nk+n-1)}+\frac{2(k+1)D_{0}r^{\beta+n+1}}{(\beta+n+1)(\beta+nk+n)}
\end{equation}
Therefore the solution of the Einstein equations with matter
field, electro-magnetic field and scalar field given by equation
(\ref{collapse2.2}) can be written as
\begin{eqnarray*}
ds^{2}=-\left[1-\frac{f(v)}{r^{n-1}}+\frac{g(v)}{(nk-1)r^{n(k+1)-2}}+\frac{e^{2}(v)}{4\pi
n(n-1)r^{2(n-1)}}-\frac{2\Lambda r^{2}}{n(n+1)} \right.
\end{eqnarray*}
\begin{equation}\label{collapse2.23}
~~~~~~~~~~~~~~\left.
-\frac{2(k+1)E(v)r}{n(nk+n-1)}-\frac{2(k+1)D_{0}r^{\beta+2}}{(\beta+n+1)(\beta+nk+n)}\right]dv^{2}+2dvdr+r^{2}d\Omega_{n}^{2}
\end{equation}
This is termed as generalization of Husain solution in higher
dimension.

We shall discuss the existence of NS in generalized Vaidya
space-time by studying radial null geodesics. In fact, we shall
examine whether it is possible to have outgoing radial null
geodesics which were terminated in the past at the central
singularity $r=0$. The nature of the singularity (NS or BH) can be
characterized by the existence of radial null geodesics emerging
from the singularity. The singularity is at least locally naked if
there exist such
geodesics and if no such geodesics exist it is a BH.

The equation for outgoing radial null geodesics can be obtained
from equation (\ref{collapse2.1}) by putting $ds^{2}=0$ and $d\Omega_{n}^{2}=0$ as
\begin{equation}\label{collapse2.24}
\frac{dv}{dr}=\frac{2}{\left(1-\frac{m(v,r)}{r^{n-1}}\right)}.
\end{equation}
It can be seen easily that $r=0,~v=0$ corresponds to a singularity
of the above differential equation. Suppose $X=\frac{v}{r}$ then
we shall study the limiting behavior of the function $X$ as we
approach the singularity at $r=0,~v=0$ along the radial null
geodesic. If we denote the limiting value by $X_{0}$ then
\begin{eqnarray}\label{collapse2.25}
\begin{array}{c}
X_{0}\\\\
{}
\end{array}
\begin{array}{c}
=lim~~ X \\
\begin{tiny}v\rightarrow 0\end{tiny}\\
\begin{tiny}r\rightarrow 0\end{tiny}
\end{array}
\begin{array}{c}
=lim~~ \frac{v}{r} \\
\begin{tiny}v\rightarrow 0\end{tiny}\\
\begin{tiny}r\rightarrow 0\end{tiny}
\end{array}
\begin{array}{c}
=lim~~ \frac{dv}{dr} \\
\begin{tiny}v\rightarrow 0\end{tiny}\\
\begin{tiny}r\rightarrow 0\end{tiny}
\end{array}
\begin{array}{c}
=lim~~ \frac{2}{\left(1-\frac{m(v,r)}{r^{n-1}}\right)} \\
\begin{tiny}v\rightarrow 0\end{tiny}~~~~~~~~~~~~\\
\begin{tiny}r\rightarrow 0\end{tiny}~~~~~~~~~~~~
 {}
\end{array}
\end{eqnarray}
Using (\ref{collapse2.22}) and (\ref{collapse2.25}), we have
\begin{eqnarray*}
\frac{2}{X_{0}}=
\begin{array}llim\\
\begin{tiny}v\rightarrow 0\end{tiny}\\
\begin{tiny}r\rightarrow 0\end{tiny}
\end{array}\left[1-\frac{f(v)}{r^{n-1}}+\frac{g(v)}{(nk-1)r^{n(k+1)-2}}+\frac{e^{2}(v)}{4\pi
n(n-1)r^{2(n-1)}}-\frac{2\Lambda r^{2}}{n(n+1)} \right.
\end{eqnarray*}
\begin{equation}\label{collapse2.26}
\left.
-\frac{2(k+1)E(v)r}{n(nk+n-1)}-\frac{2(k+1)D_{0}r^{\beta+2}}{(\beta+n+1)(\beta+nk+n)}\right]
\end{equation}
Now choosing, $f(v)=f_{0}v^{n-1}$, $g(v)=g_{0}v^{n(k+1)-2}$,
$e(v)=e_{0}v^{n-1}$ and $E(v)=\frac{E_{0}}{v}$, we obtain the
algebraic equation of $X_{0}$ as
\begin{equation}\label{collapse2.27}
f_{0}X_{0}^{n}-\frac{g_{0}}{nk-1}~X_{0}^{nk+n-1}-\frac{e_{0}^{2}}{4\pi
n(n-1)}~X_{0}^{2n-1} +\frac{2(k+1)E_{0}}{n(nk+n-1)}  -X_{0}+2=0
\end{equation}
From above equation, we see that the equation is completely
independent of $\Lambda$ and $D_{0}$.\\
\begin{figure}\label{table}
{\bf Table:} The table represents the values of $X_{0}$ for
different
values of parameters in various dimensions.\\\\
\includegraphics[height=4.5in, width=6.7in]{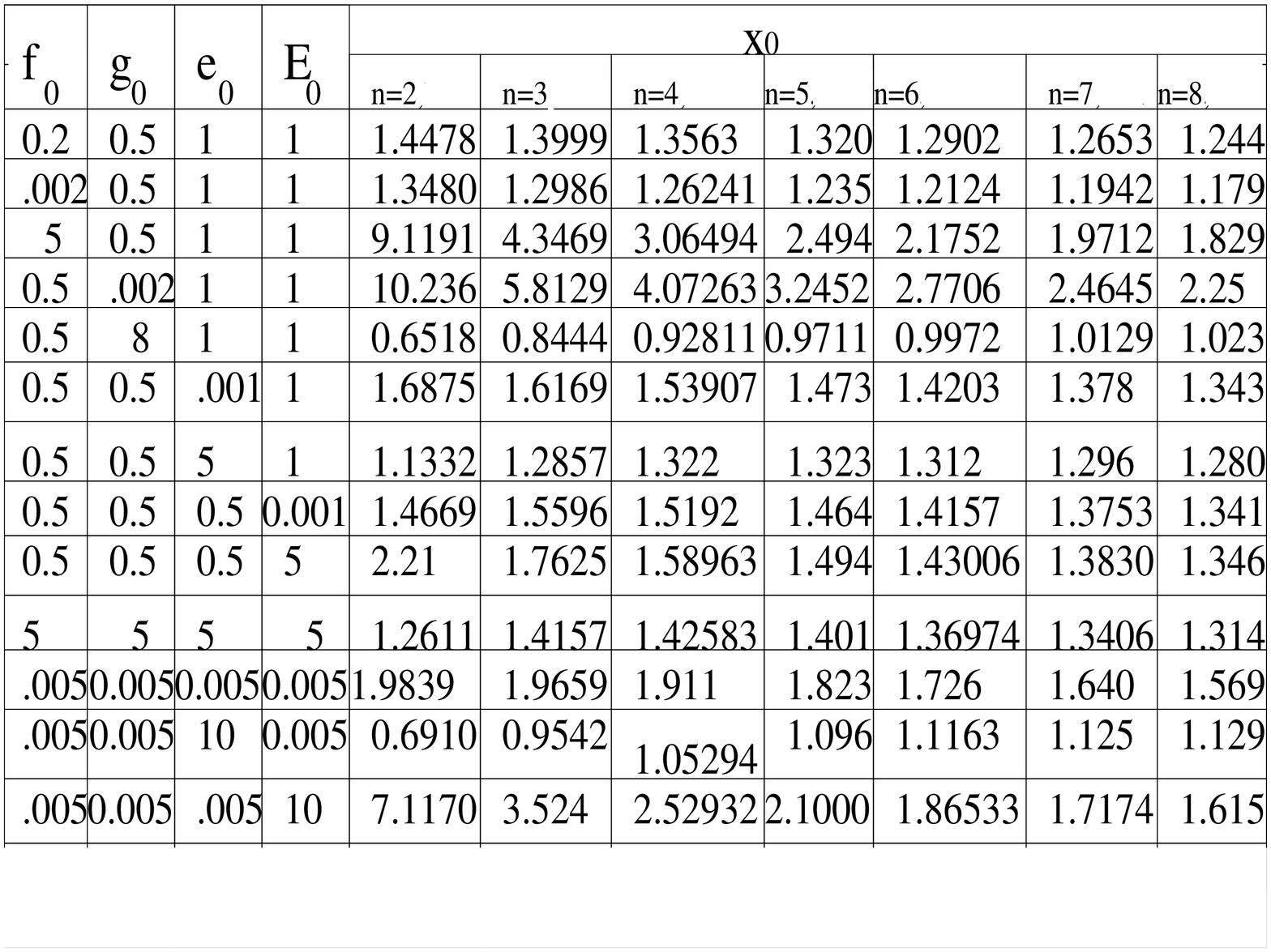}~~~~~~~~

\end{figure}
Now if we get only non positive solution of the equation we can
assure the formation of a BH. Getting a positive root indicates a
chance to get a NS. Since the obtained equation is an highly
complicated one, it is extremely difficult to find out an analytic
solution of $X_{0}$ in terms of the variables involved. So our
idea is to find out different numerical solutions of $X_{0}$, by
assigning particular numerical values to the associated variables.

The different solutions of $X_{0}$ for different sets of
parametric values ($f_{0},g_{0},e_{0},E_{0}$) are given here in a
tabular form. It is seen that over a large range of parametric
values the positive solution for $X_{0}$ is the only outcome,
which immediately implies the occurrence of NS as the outcome of
the collapse.
\section{Impact of Accretion upon the collapsing body}\label{chap2}
The highly massive star, white dwarf or the other things which are
going to collapse to form a singularity they do so as their mass
limit crosses the Chandrasekhar's limit. In general cases it
happens as the object inhells mass from the dust cloud nearby or
from a companion star which is in a binary system with the
concerned object. Collapse and accretion together has been studied
in literature since a long time(Wagh, S.M. 2002; Vorobyov,E.I. and
Basu,S. 2005; Johnson,J.L. 2010).

Now if we assume that the object we are concerning is collapsing
with an accretion procedure together then we can modify our mass
function to the expression given by
\begin{equation}\label{collapse2.28}
M-\dot{M}dv=M-\Omega_{n}r^{n}T_{0}^{1}dv
\end{equation}
The $T_{0}^{1}$ represents the non-diagonal stress energy tensor
component. The explicit form of $T_{0}^{1}$ is given by the
equations (\ref{collapse2.3}),(\ref{collapse2.4}) and
(\ref{collapse2.5}). Using which we can calculate the value of
$T_{0}^{1}$ and the modified form of the $f(r, ~v)$ represented by
a $\bar{f}(r, ~v)$ has the expression
\begin{equation}\label{collapse2.29}
\bar{f}(r, ~v)=f(r,~v)+\frac{\Omega_{n}\sigma}{r^{n}}
\end{equation}
From which with the assumption $\sigma=\sigma_{0}v$ the equation of $X_{0}$ comes as
\begin{equation}\label{collapse2.30}
f_{0}X_{0}^{n}-\frac{g_{0}}{nk-1}~X_{0}^{nk+n-1}-\frac{e_{0}^{2}}{4\pi
n(n-1)}~X_{0}^{2n-1}-\frac{\pi^{\frac{n}{2}}\sigma_{0}}{\Gamma(1+\frac{n}{2})}X_{0}^{2} +\frac{2(k+1)E_{0}}{n(nk+n-1)}  -X_{0}+2=0
\end{equation}

\begin{figure}\label{figure}
\includegraphics[height=2in, width=2in]{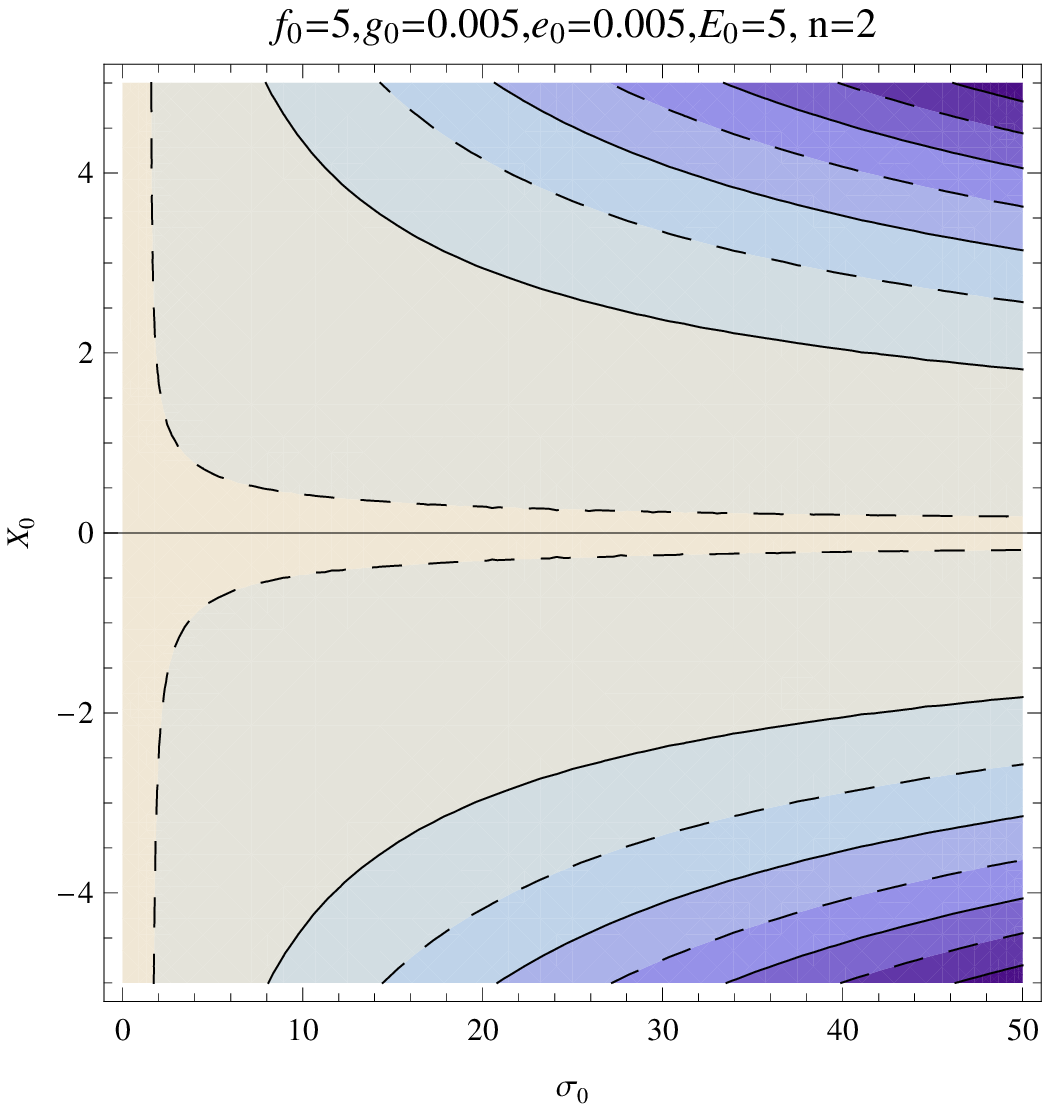}~~\includegraphics[height=2in, width=2in]{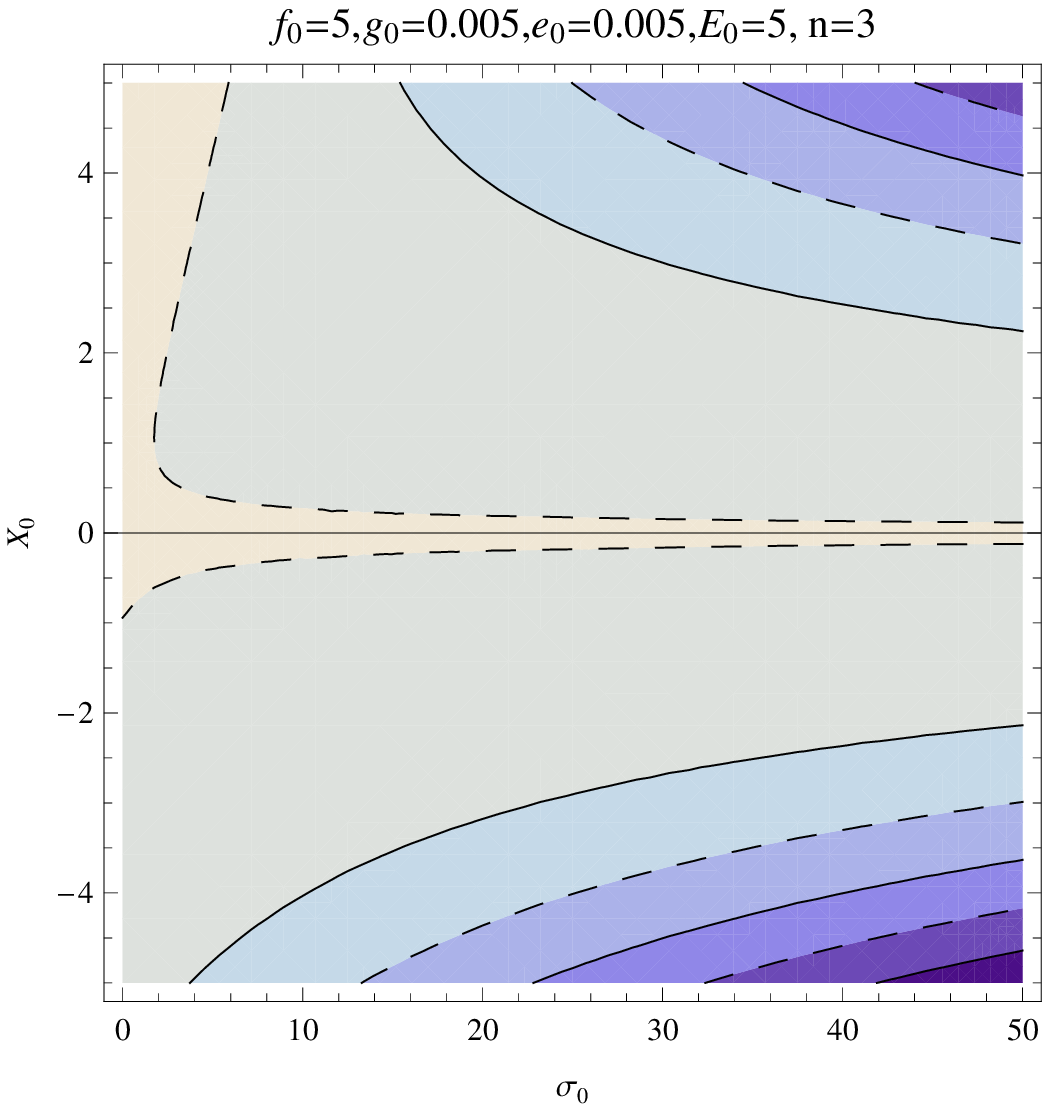}~~\includegraphics[height=2in, width=2in]{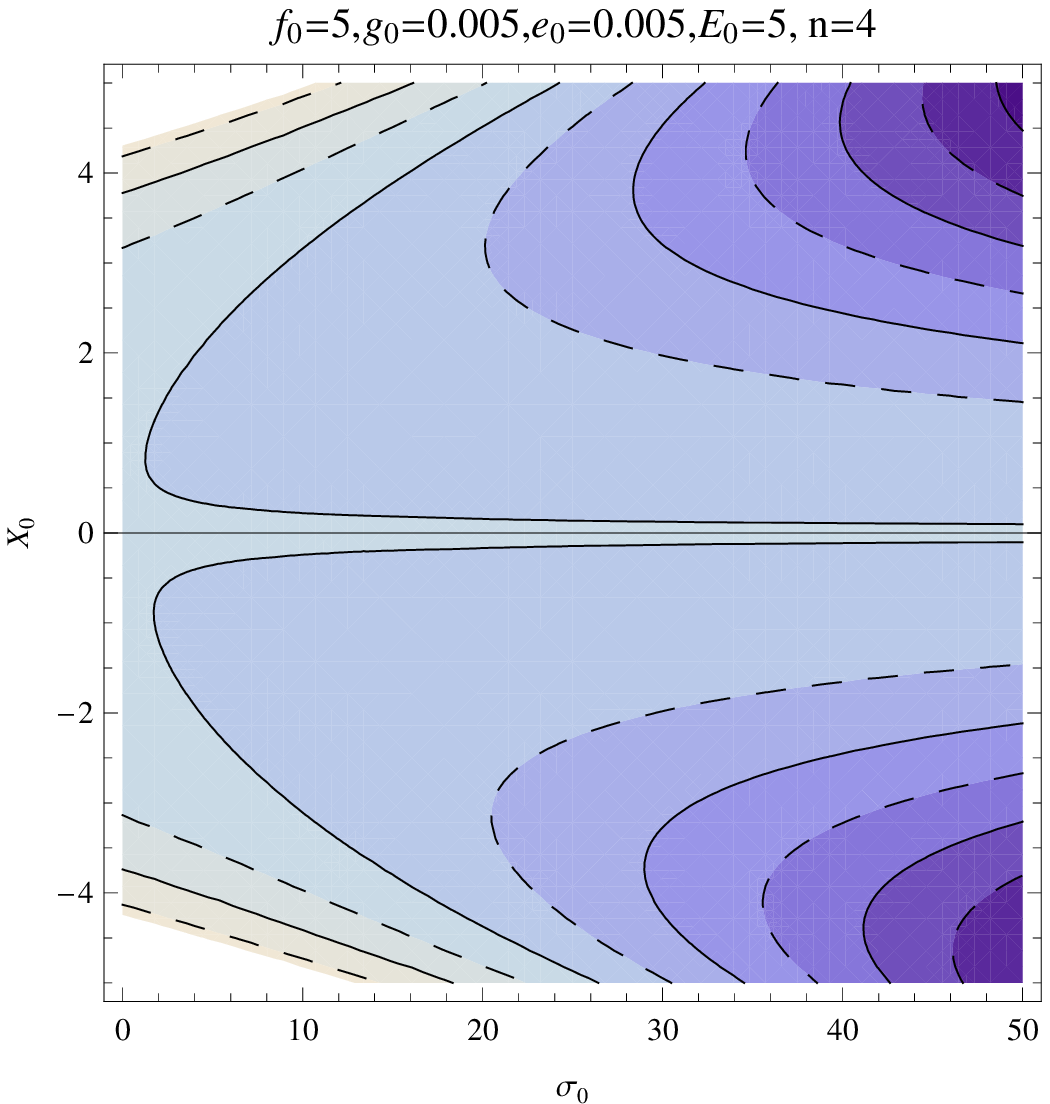}\\
\includegraphics[height=2in, width=2in]{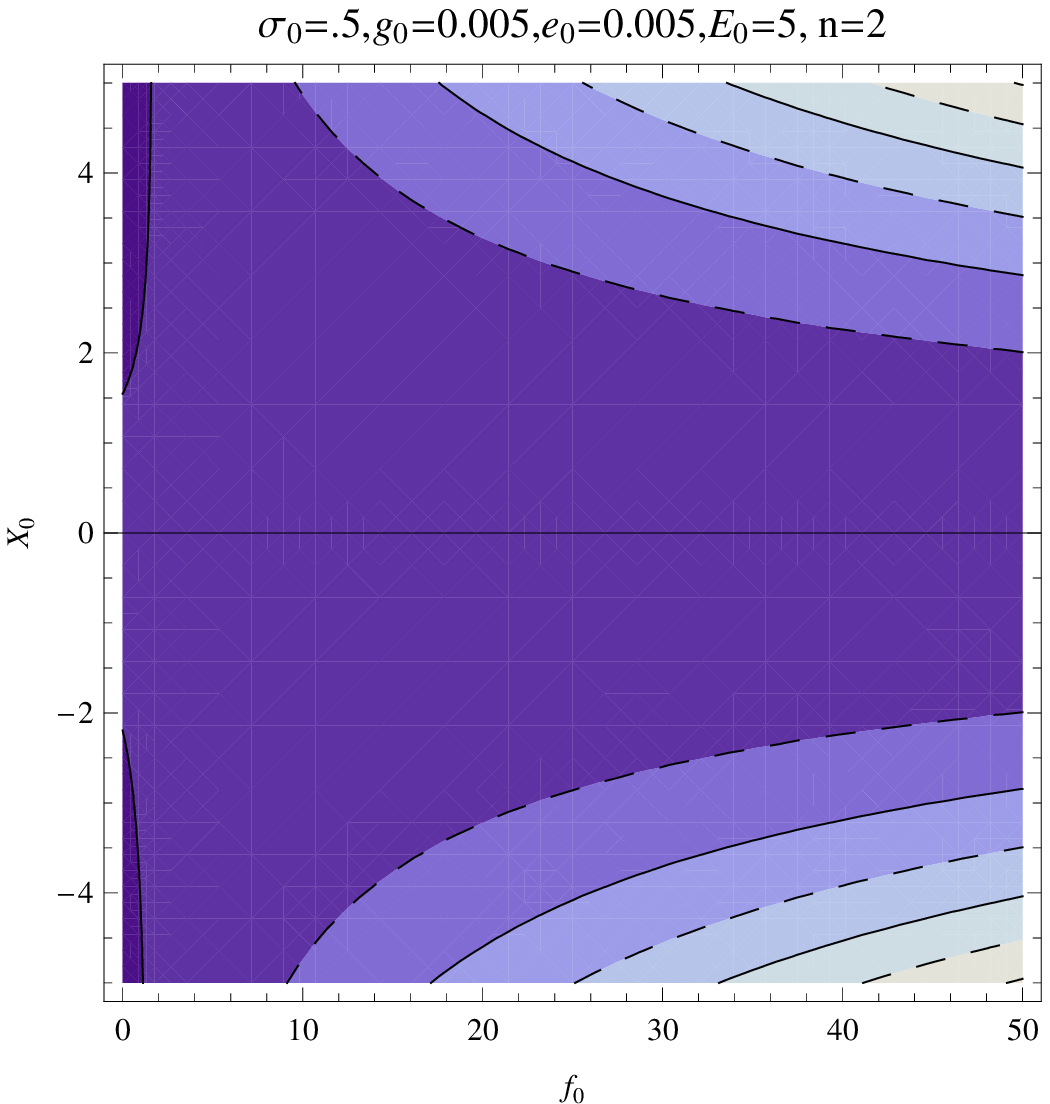}~~\includegraphics[height=2in, width=2in]{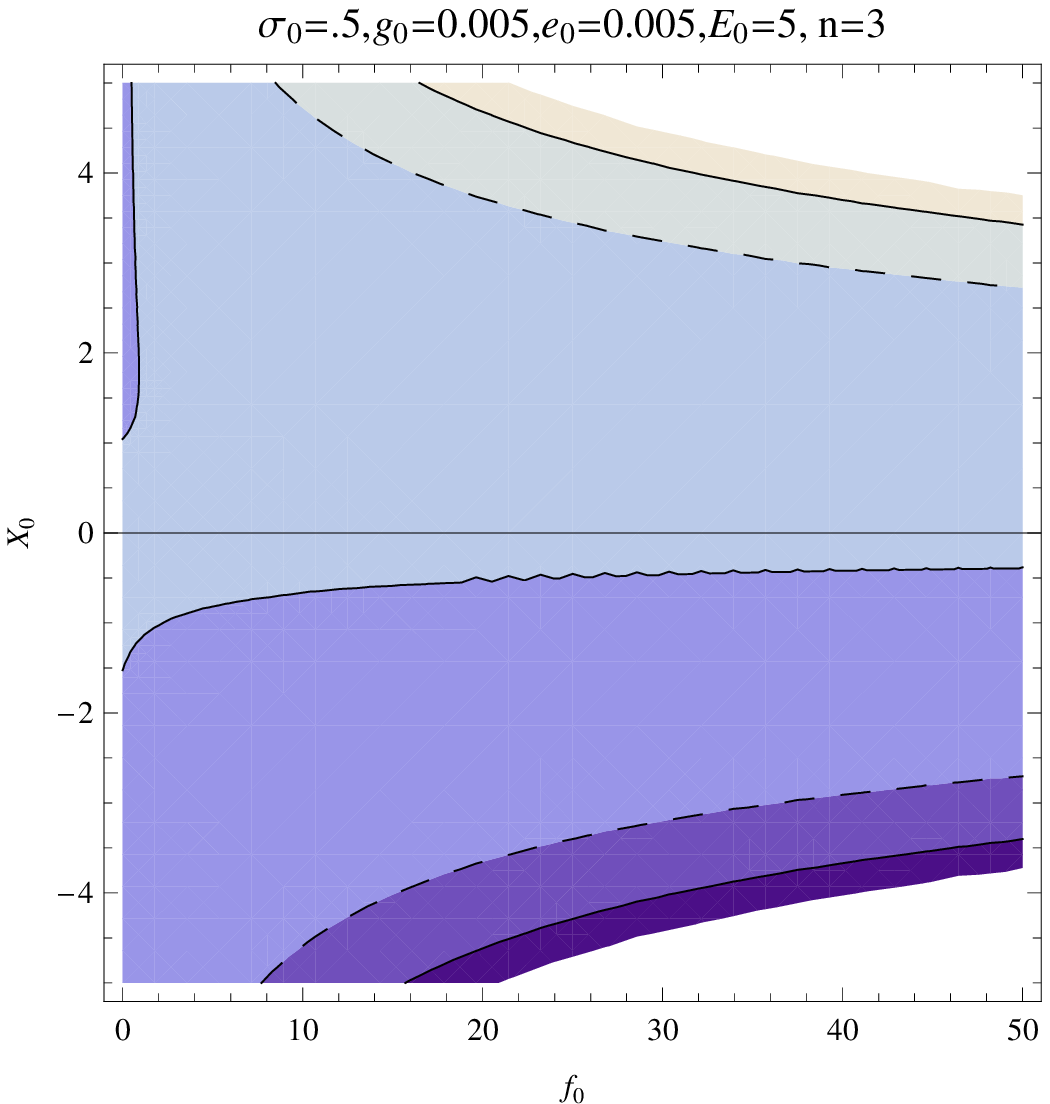}~~\includegraphics[height=2in, width=2in]{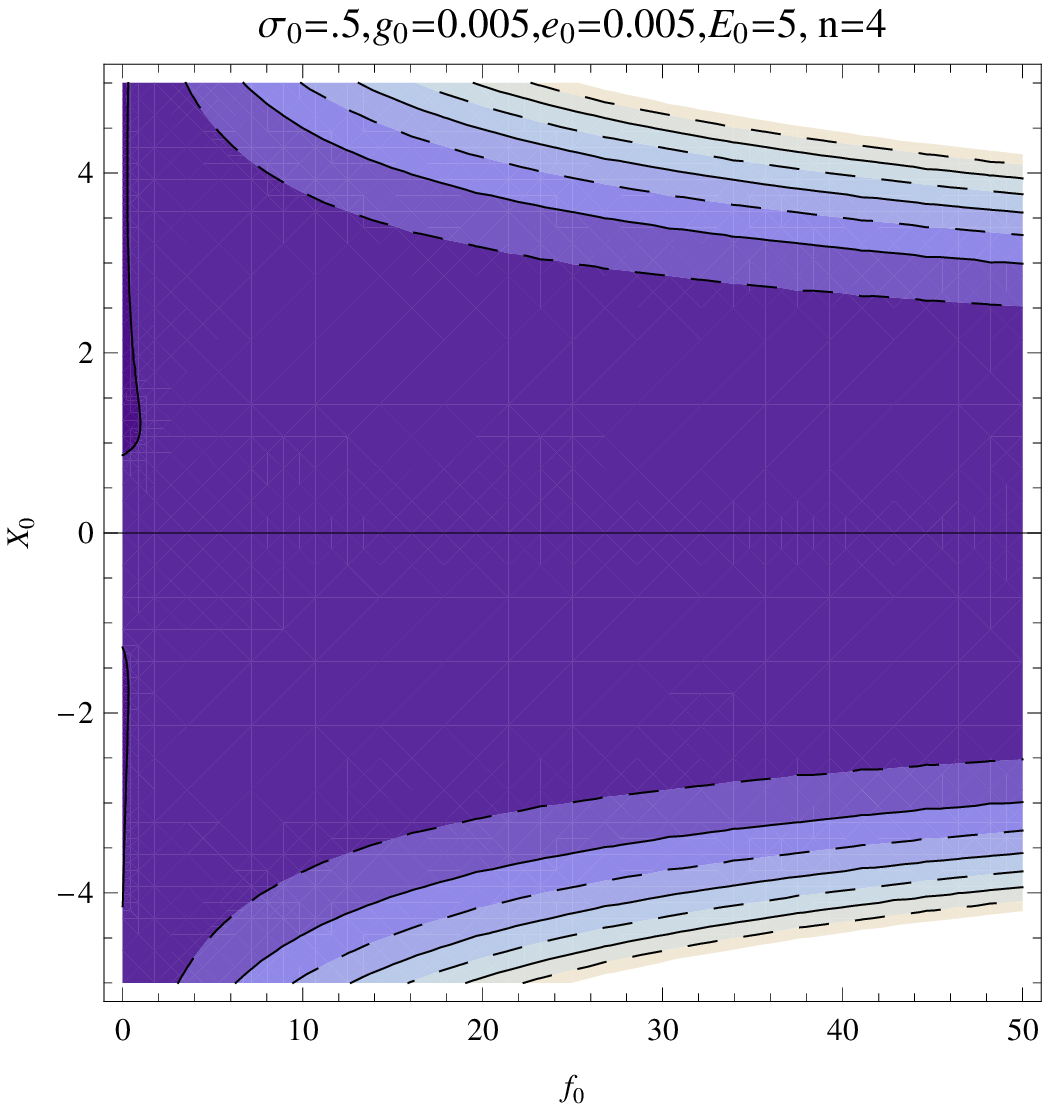}\\
\includegraphics[height=2in, width=2in]{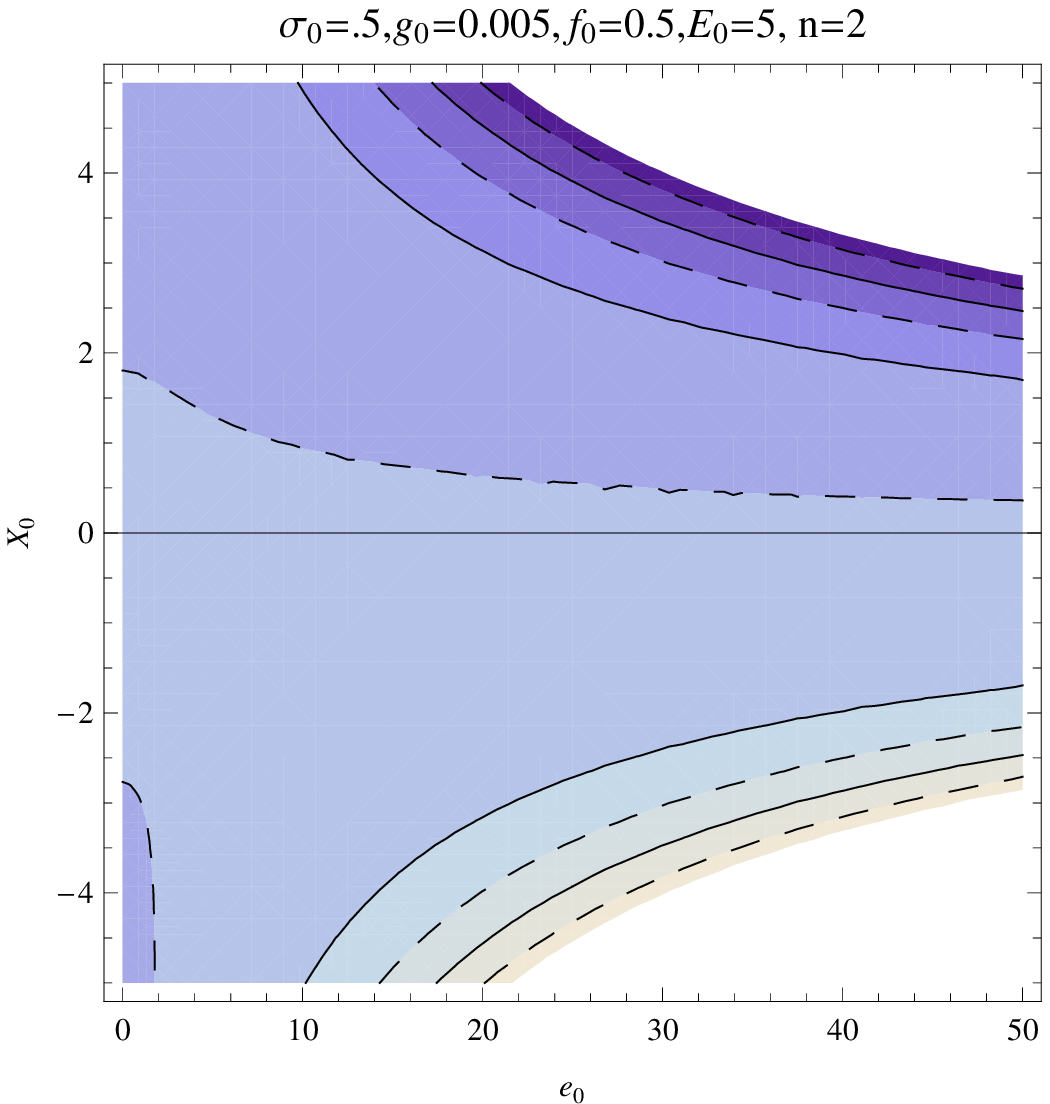}~~\includegraphics[height=2in, width=2in]{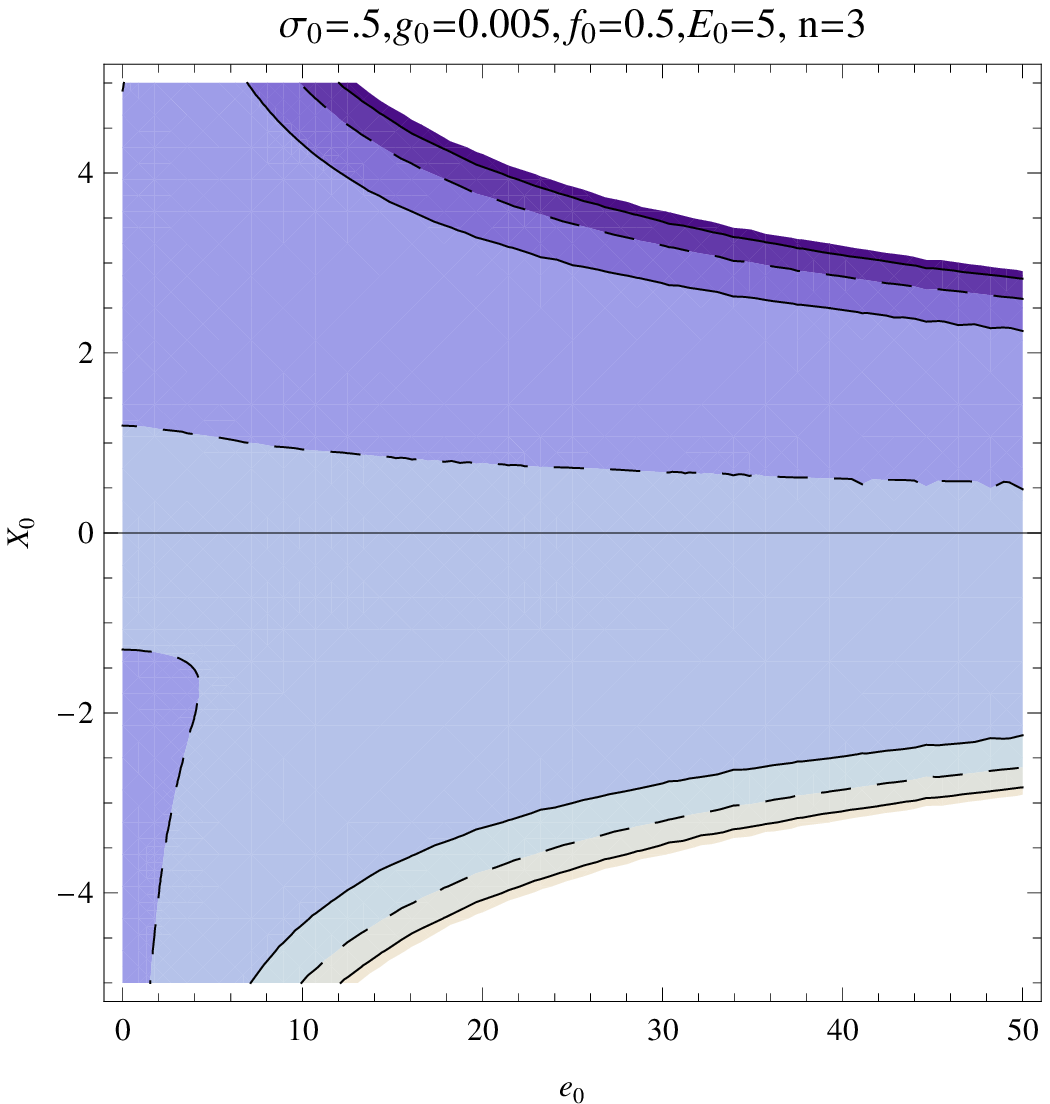}~~\includegraphics[height=2in, width=2in]{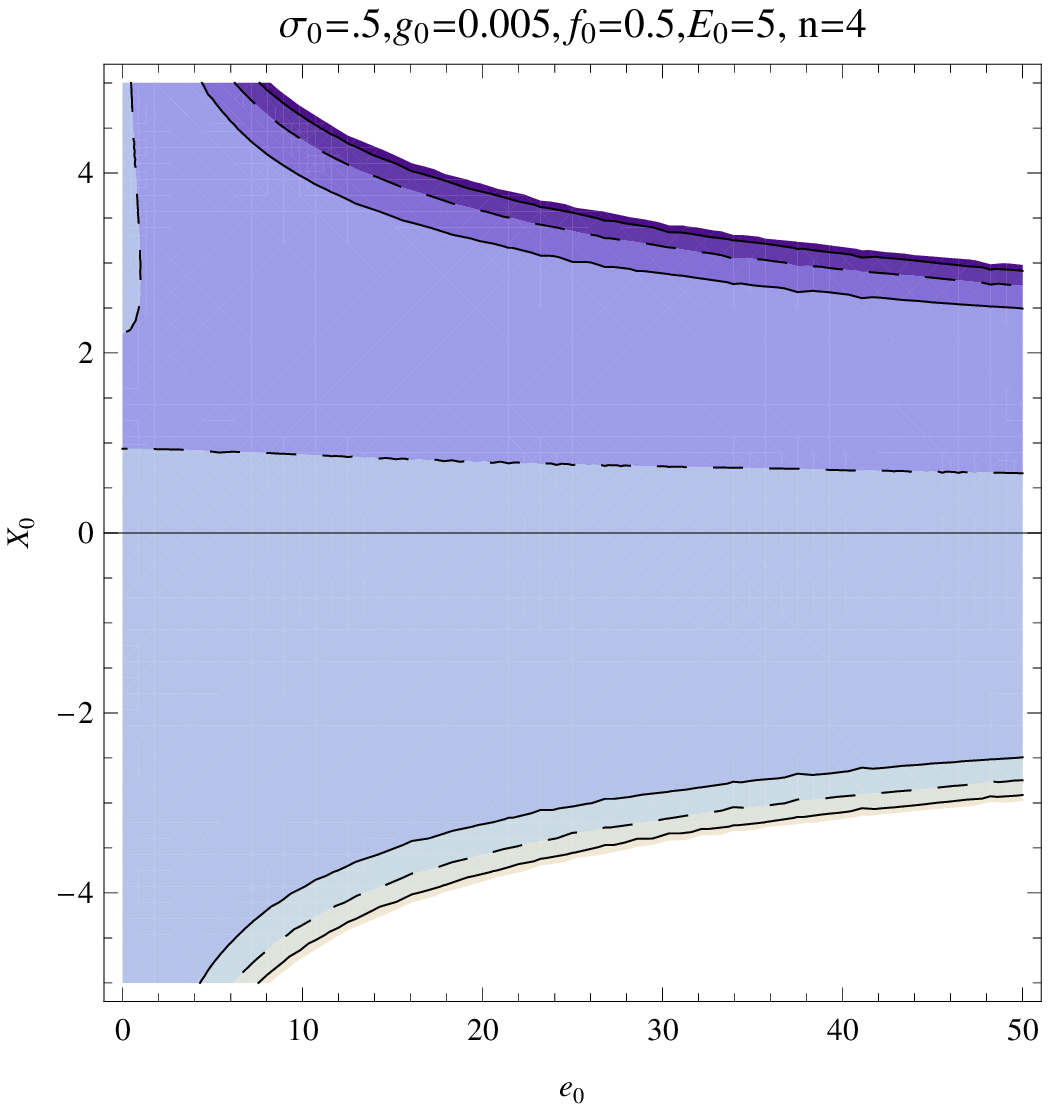}\\
The figures represent the plot of $X_{0}$ (\ref{collapse2.30})
with the variation of different parameters in various dimensions
when some fluids are accreting upon collapsing objects.
\end{figure}
Here again the analytical solution of the equation is bit tough. Rather we can solve $X_{0}$
easily for different parametric values. But after going through this procedure we found that
$X_{0}$ does not possess any values in the negative region . Hence this result derives us into
plotting graphs for $X_{0}$, in various regions in order to get an explicit idea of the nature
of solution of the concerned equation. After plotting $X_{0}$ with different ranges of all other
parameters it is quite clear from the graphs that $X_{0}$ always has a positive value for any
combination of the other parameters. So even if accretion procedure is going upon the collapsing
object the chance of having NS always will be there.

\begin{figure}\label{figure}
\includegraphics[height=2in, width=2in]{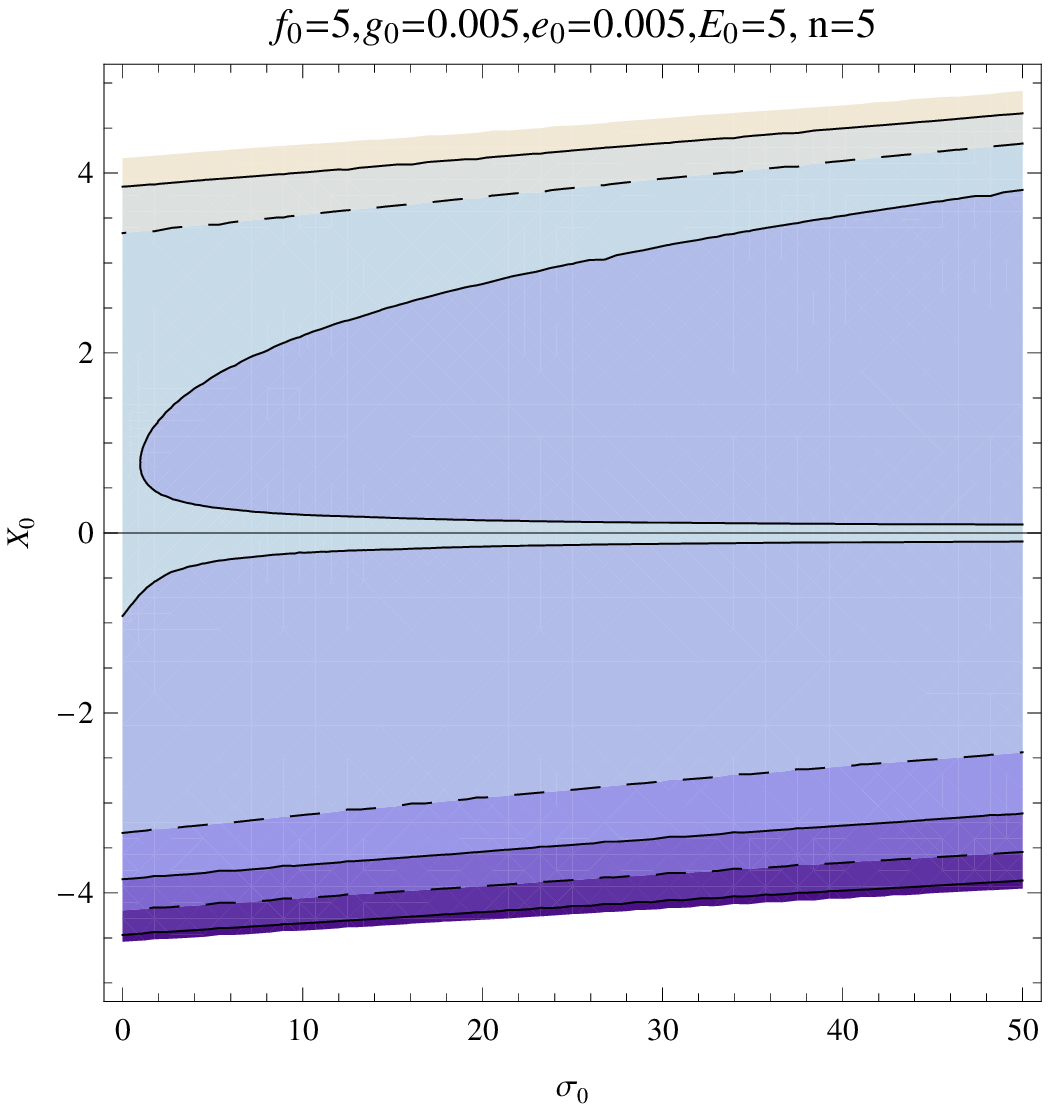}~~\includegraphics[height=2in, width=2in]{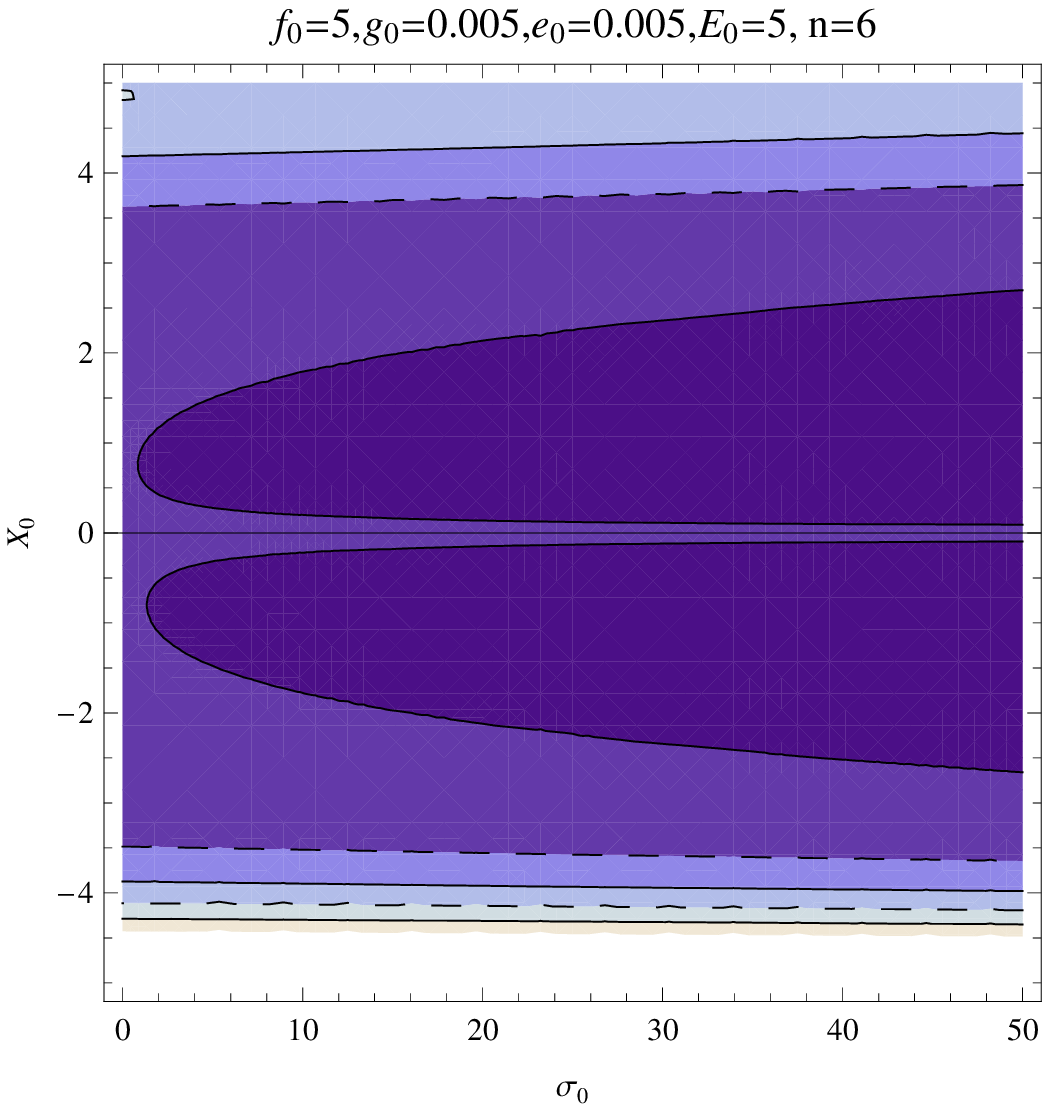}~~\includegraphics[height=2in, width=2in]{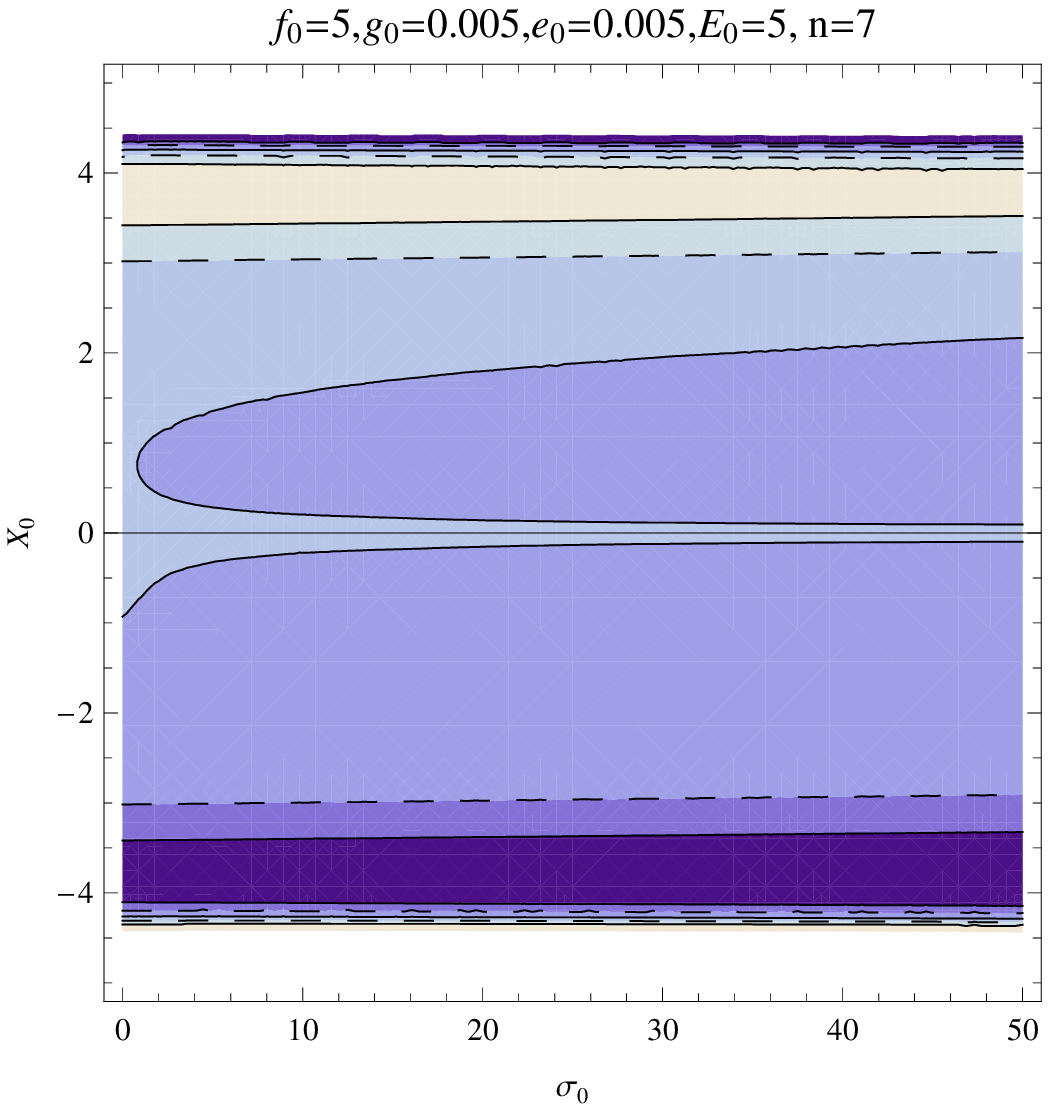}\\
\includegraphics[height=2in, width=2in]{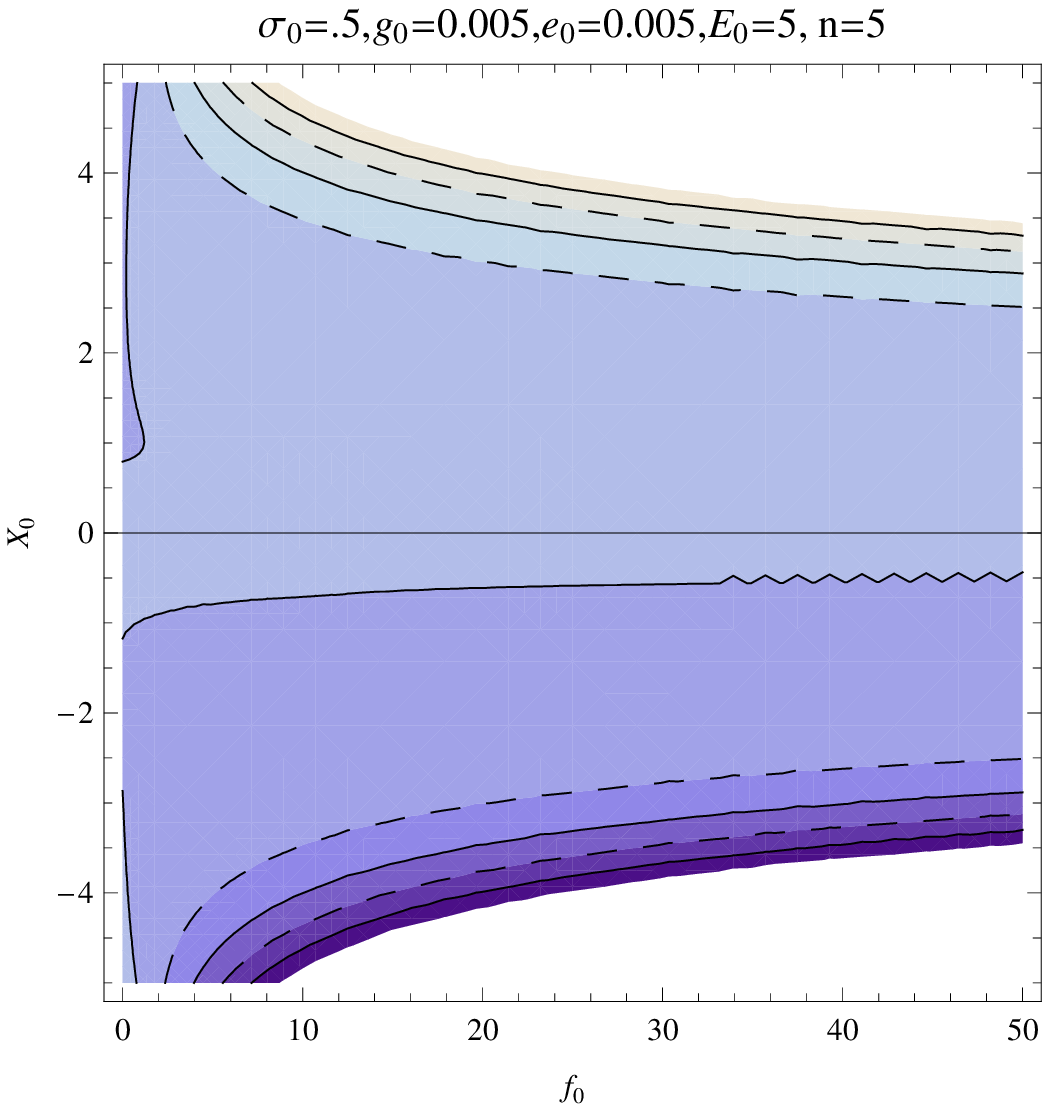}~~\includegraphics[height=2in, width=2in]{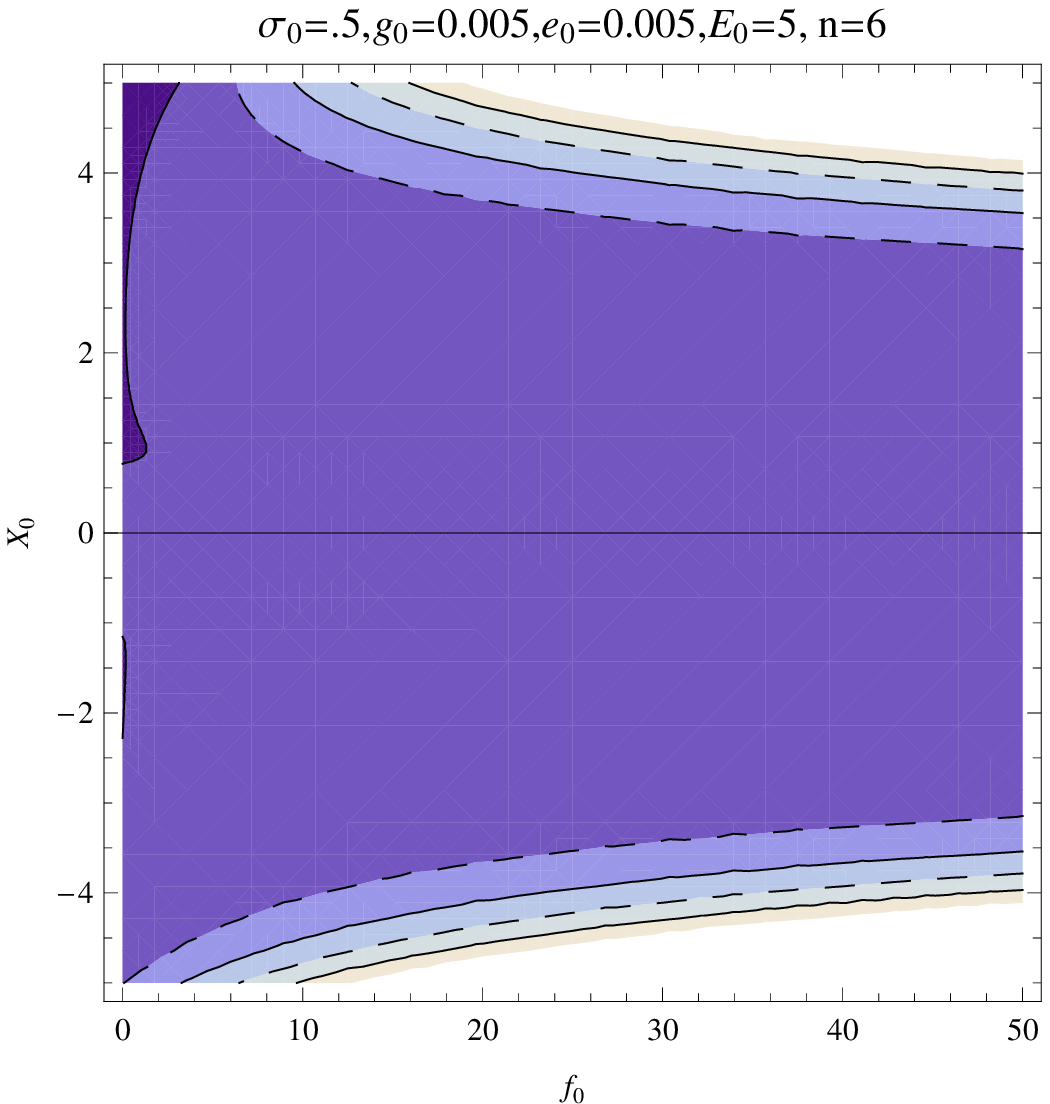}~~\includegraphics[height=2in, width=2in]{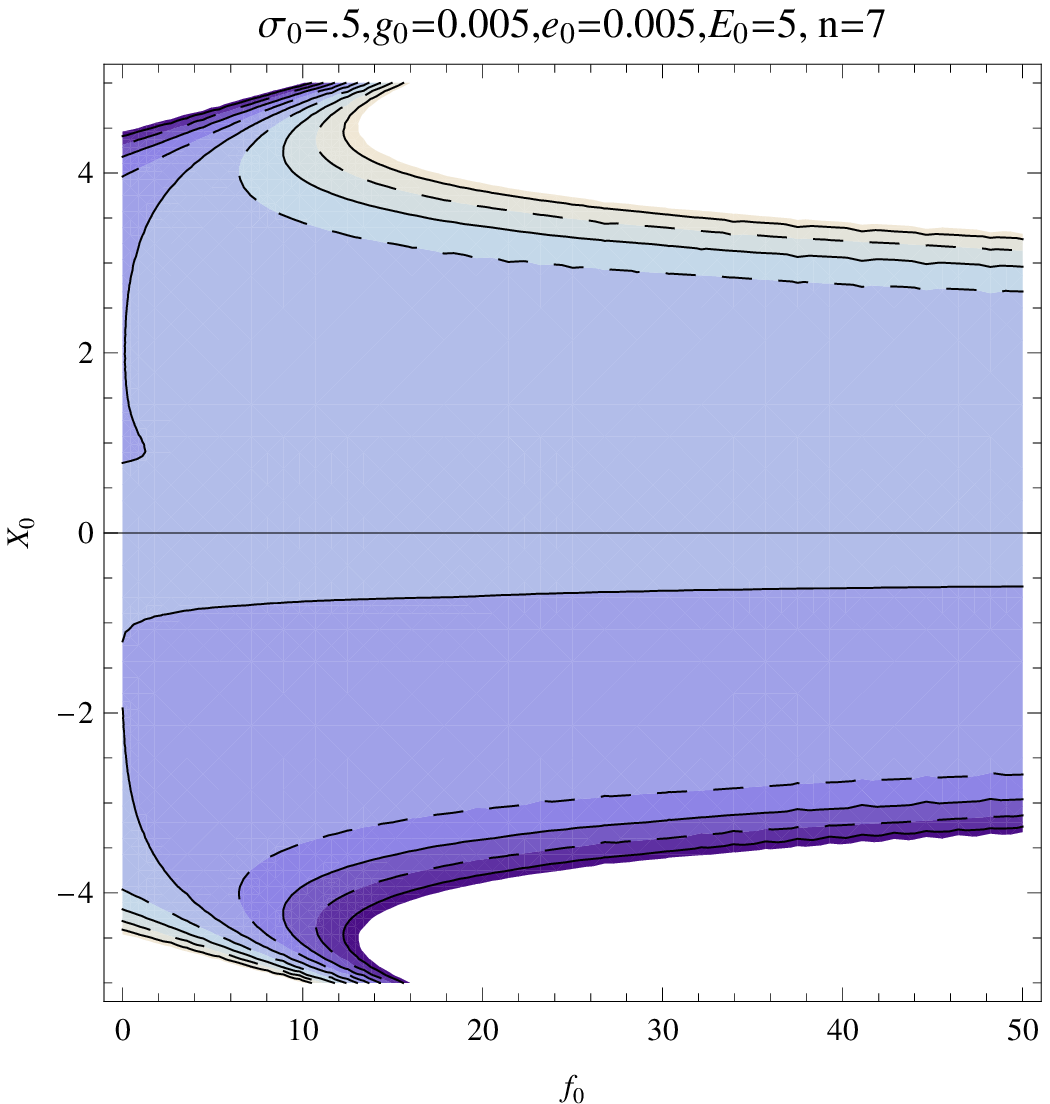}\\
\includegraphics[height=2in, width=2in]{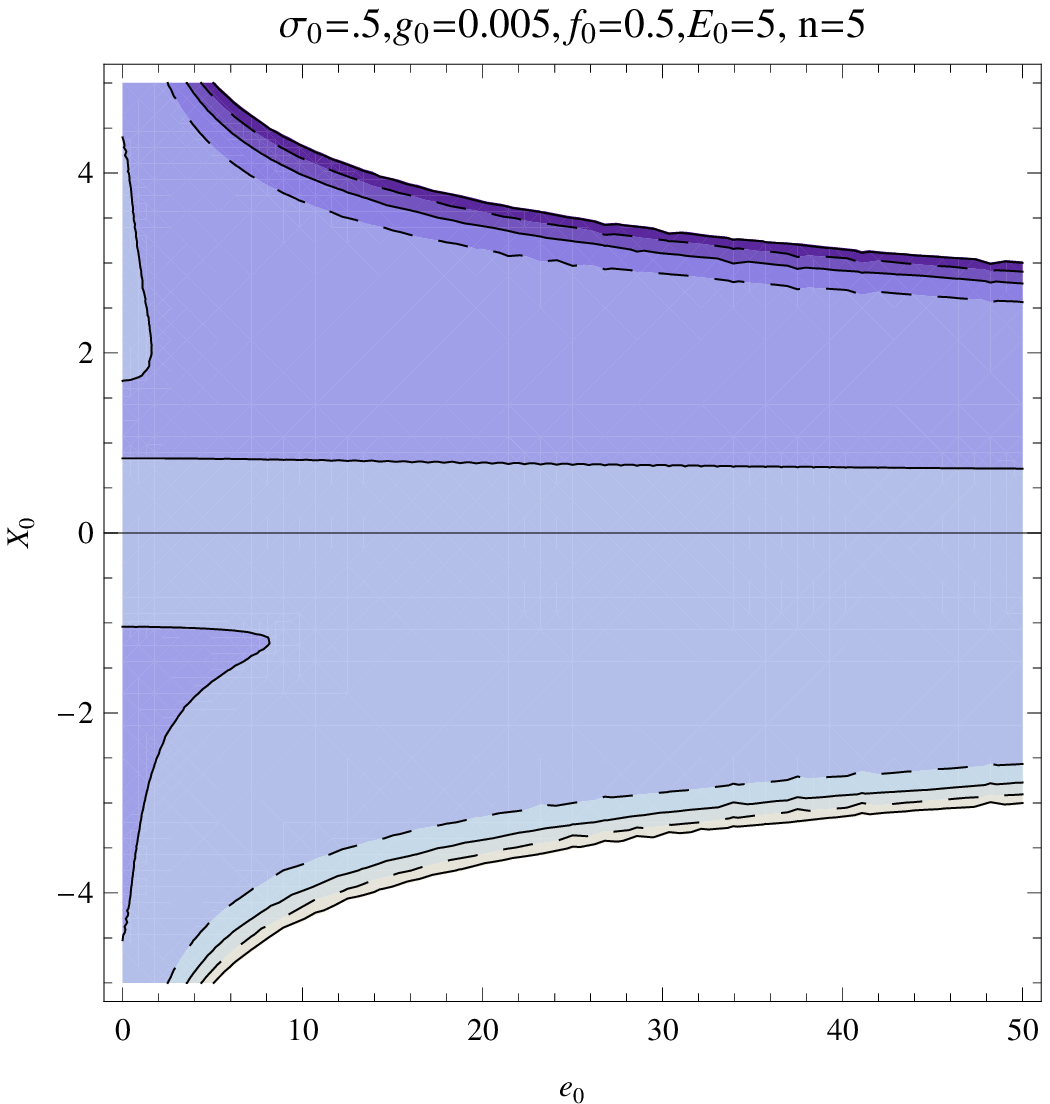}~~\includegraphics[height=2in, width=2in]{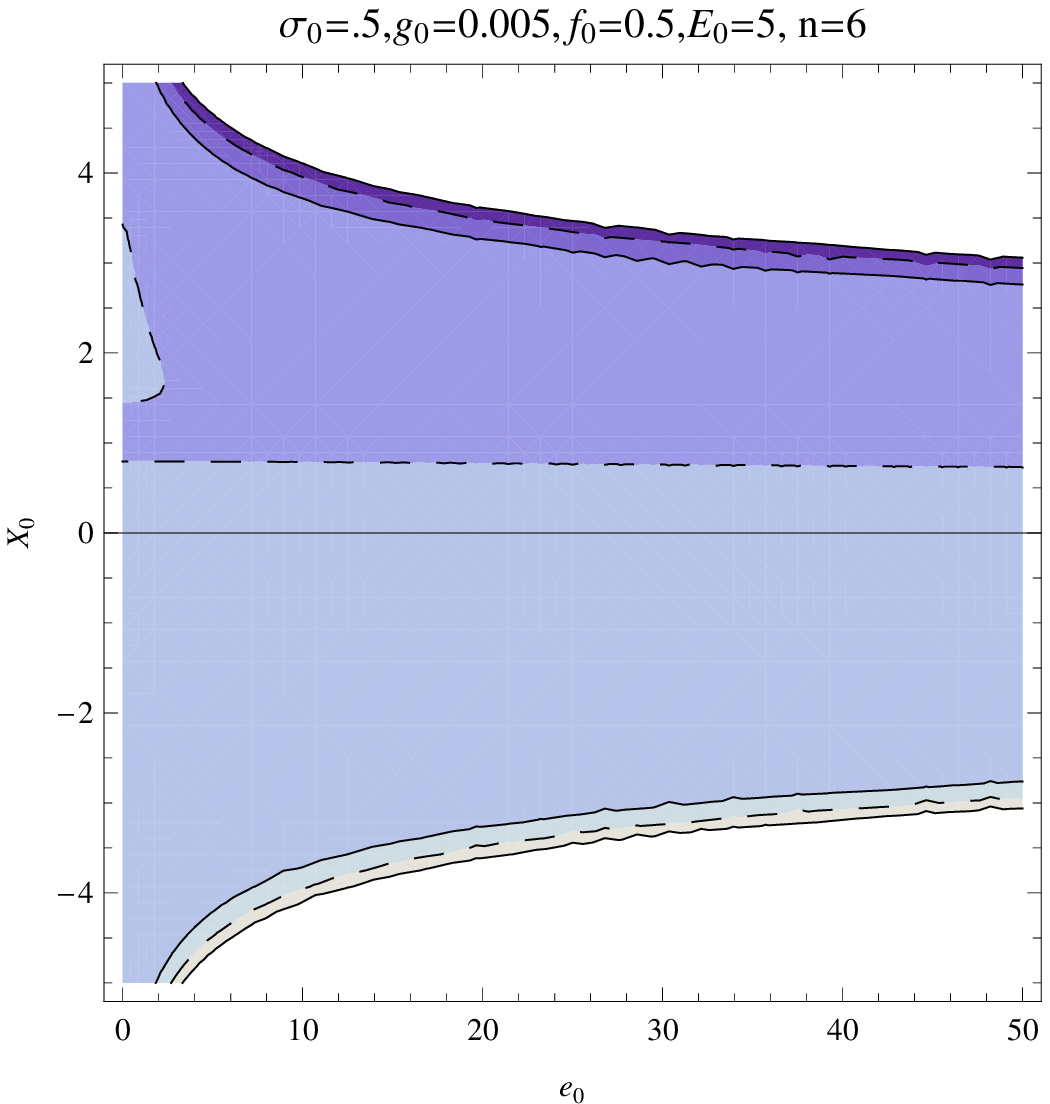}~~\includegraphics[height=2in, width=2in]{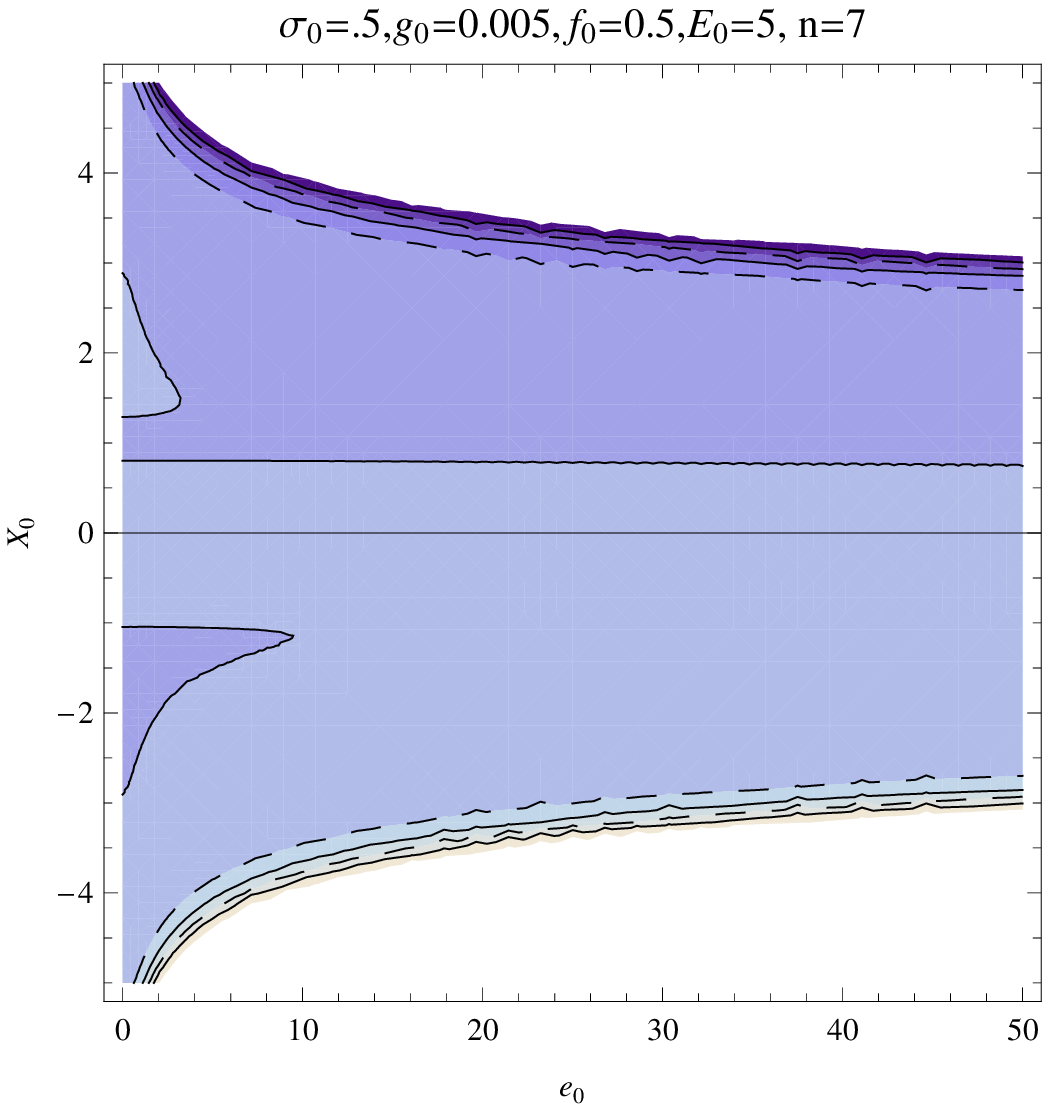}\\
The figures represent the plot of $X_{0}$ (\ref{collapse2.30})
with the variation of different parameters in various dimensions.
\end{figure}

\section{Conclusion}\label{chap3}

Here we have considered the $(n+2)$-dimensional Husain space-time
in presence of radiating null fluid, barotropic fluid, an
electro-magnetic field and a scalar field with potential. The
corresponding solutions have been obtained. Next we have analyzed
the gravitational collapse of a star in Husain space-time. The
existence of outgoing radial null geodesic has been thoroughly
investigated in order to characterize the nature of central
singularity formed as the end state of collapse. Equation of
radial null geodesic is formed and the values of the variable
$X_{0}$ is computed for different values of parameters in
different dimensions. Surprisingly in all the cases it was found
that irrespective of the values of the parameters chosen, the
value of $X_{0}$ is always positive and hence the singularity
formed is always a naked singularity. This has been shown in a
tabular form. Impact of accretion phenomenon on the gravitational
collapse of a star has also been taken into account. The mass
function for a collapsing procedure is considered and from it, the
equation of $X_{0}$ is constructed for various dimensions. To
study this case graphs were drawn for different values of the
parameters ($f_{0},g_{0},e_{0},E_{0}$) and it was found that the
figures did not show any tendency of sneaking into the negative
region. Hence black holes may not be the correct option for Husain
metric in any dimension. Formation of event horizons is totally
barred in Husain metric. So the only possibility is a naked
singularity. Hence this is a significant
counter example of the cosmic censorship hypothesis.\\\\\\

{\bf Acknowledgement :}\\
\\
RB thanks West Bengal State Govt. for
awarding SRF.

\end{document}